\documentclass[aps,prd,showpacs,notitlepage,nofootinbib,superscriptaddress,floatfix,showkeys,twocolumn]{revtex4-1}
\pdfoutput=1
\usepackage{array}
\usepackage[normalem]{ulem} 
\usepackage{verbatim}
\usepackage{graphicx}
\usepackage{amsmath}
\usepackage{amsfonts}
\usepackage{amssymb}
\usepackage{empheq}
\usepackage{xcolor, soul}
\usepackage{epstopdf}
\usepackage{float}
\usepackage{subfigure}
\usepackage{caption}
\usepackage{subcaption}
\usepackage{soul}
\usepackage{placeins}
\usepackage[mathscr]{eucal}
\usepackage{hyperref}
\usepackage{ragged2e}
\hypersetup{
     colorlinks   = true,
     citecolor    = red,
     linkcolor    = blue,
     urlcolor     = blue,
}
\definecolor{ao(english)}{rgb}{0.0, 0.5, 0.0}

\newcommand{\noi}{\noindent}
\newcommand{\beq}{\begin{equation}}
\newcommand{\eeq}{\end{equation}}
\newcommand{\bea}{\begin{eqnarray}}
\newcommand{\eea}{\end{eqnarray}}

\newcommand{\Mp}{M_P}
\newcommand{\Tbh}{T_{\rm BH}}
\newcommand{\Mbh}{M_{\rm BH}}

\newcommand{\Trh}{T_{\rm RH}}
\newcommand{\Hrh}{H_{\rm RH}}
\newcommand{\arh}{a_{\rm RH}}

\newcommand{\Mrh}{M_{\rm RH}}

\newcommand{\Min}{M_{\rm in}}
\newcommand{\Hin}{H_{\rm in}}

\newcommand{\ain}{a_{\rm in}}
\newcommand{\xin}{x_{\rm in}}
\newcommand{\rhoin}{\rho_{\rm in}}

\newcommand{\rhorh}{\rho_{\rm RH}}

\newcommand{\Rrh}{R_{\rm RH}}

\newcommand{\gameff}{\gamma_{\rm eff}}
\newcommand{\gamc}{\gamma_{\rm c}}

\graphicspath{{./figs/}}
\begin{document}


\title{PBH runaway during reheating}

\author{Md Riajul Haque}%
\email{riaj1994@sjtu.edu.cn}
\affiliation{\,Tsung-Dao Lee Institute \& School of Physics and Astronomy, Shanghai Jiao Tong University,
Shanghai 201210, China}.
\author{Mathieu Gross}
\email{mathieu.gross@ijclab.in2p3.fr}
\affiliation{
	Universit\'e Paris-Saclay, CNRS/IN2P3, IJCLab, 91405 Orsay, France 
}
\author{Mathieu Houlier}
\email{mathieu.houlier@universite-paris-saclay.fr}
\affiliation{
	Universit\'e Paris-Saclay, CNRS/IN2P3, IJCLab, 91405 Orsay, France 
}
\author{Yann Mambrini}
\email{yann.mambrini@ijclab.in2p3.fr}
\affiliation{
	Universit\'e Paris-Saclay, CNRS/IN2P3, IJCLab, 91405 Orsay, France 
}

\begin{abstract}

The growth of primordial black holes through the absorption of the surrounding 
plasma has recently been shown to exhibit a critical behavior during radiation 
domination. We extend this analysis to the reheating era and derive analytical 
solutions for general reheating histories. We show that reheating modifies the 
critical condition for runaway absorption, making it dependent on both the 
reheating dynamics and the black-hole formation time. We identify two distinct 
regimes: runaway growth occurring \emph{during} reheating, or \emph{being triggered} after the 
onset of radiation domination by the mass accumulated during reheating. More 
generally, we derive a simple composition law describing how independent mass-
growth mechanisms combine across successive cosmological eras. Applying it to 
radiation absorption and inflaton accretion, we obtain analytical results in 
excellent agreement with the full numerical evolution.
\end{abstract}

\maketitle
\twocolumngrid

\section{Introduction}

The evolution of black-hole masses through accretion has been studied for
decades, beginning with the pioneering analyses of spherical accretion by
Bondi~\cite{Bondi:1952ni} and its relativistic extension by
Michel~\cite{Michel:1972oeq}, together with early investigations of black-hole
evolution in cosmological backgrounds by Novikov and
Zel'dovich~\cite{Zeldovich:1967} and Carr and
Hawking~\cite{Carr:1974nx}. In particular, the latter showed that the
cosmological expansion strongly constrains the growth of primordial black
holes through accretion during radiation domination. More recently,
general-relativistic treatments have further emphasized the importance of
relativistic effects in determining the accretion history and mass evolution
of PBHs~\cite{Das:2025vts}.

Primordial black holes (PBHs), formed at the earliest stages of the Universe,
continuously exchange energy with their cosmological environment through
Hawking evaporation and the absorption of the surrounding plasma. The
competition between these two processes determines whether a PBH evaporates
or instead undergoes a phase of rapid mass growth. The possibility of rapid
black-hole growth in a thermal environment has also been investigated for
black holes immersed in a heat bath~\cite{Barrau:2022bfg}. Recently, it was
shown that during radiation domination this competition exhibits a remarkable
critical behavior. Primordial black holes formed above a universal collapse
efficiency,
\(
\gamma_c \simeq 0.395,
\)
enter a runaway absorption regime in which their mass grows without
bound~\cite{Haque:2026vvp}.

The existence of such a universal threshold is a direct consequence of the
scale-free nature of a radiation-dominated Universe. Since the expansion
history introduces no preferred physical scale, the critical collapse
efficiency is independent of the PBH formation time or initial mass.

The early Universe, however, is not expected to become radiation dominated
immediately after inflation. Before the onset of the standard hot Big Bang,
the Universe undergoes a reheating phase during which the oscillating inflaton
dominates the expansion while continuously producing a thermal
bath~\cite{Garcia:2023obw,Garcia:2020wiy,Garcia:2020eof}. During this epoch,
the Hubble expansion, the radiation temperature, and the horizon mass evolve
differently from their radiation-dominated counterparts. It is therefore
natural to ask whether the universal runaway condition survives in this more
general cosmological background.

In this work, we demonstrate that reheating fundamentally breaks the
universality of the critical collapse efficiency. The amount of mass absorbed
before radiation domination depends on the PBH formation time, implying that
the collapse threshold becomes a function of both the reheating dynamics and
the initial PBH mass. We derive analytical expressions for the critical
collapse efficiency for general inflaton equations of state and temperature
evolutions during reheating. We show that two physically distinct runaway
regimes exist, corresponding respectively to runaway growth during reheating
and to runaway triggered only after the onset of radiation domination.

More generally, we formulate a composition law describing how successive
cosmological eras and multiple independent mass-growth mechanisms combine to
determine the effective collapse threshold. This framework allows the
evolution of primordial black holes to be described analytically across
successive cosmological histories and provides a unified treatment of
reheating, radiation domination, and additional accretion processes.

The paper is organized as follows. In Sec.~\ref{Sec:PBHevolution}, we
study the PBH mass evolution during reheating, beginning with the
competition between absorption and Hawking evaporation and deriving the
critical conditions for runaway growth during reheating and at the onset
of radiation domination. In Sec.~\ref{Sec:generalization}, we generalize
the analysis to multiple mass-growth processes and derive the composition
law governing their combined effect, which we then apply to radiation
absorption and inflaton accretion. In Sec.~\ref{Sec:refinement}, we refine
the treatment of the reheating-to-radiation transition by retaining both
the inflaton and radiation contributions to the Hubble expansion and
compare the analytical result with the full numerical evolution. We
summarize our main results in Sec.~\ref{Sec:conclusion}. Finally, in
Appendix~\ref{app:dof}, we examine the impact of the temperature
dependence of the relativistic degrees of freedom on the critical collapse
efficiency during radiation domination.

\section{PBH Mass Evolution During Reheating}\label{Sec:PBHevolution}

\subsection{Absorption and evaporation}

In Refs. \cite{Haque:2026vvp} and \cite{Kallifatides:2026sik}, the 
authors derived the complete evolution equation for a primordial 
black hole, accounting for both Hawking evaporation and particle 
absorption from the surrounding primordial plasma:

\beq
   \frac{d\Mbh}{dt} = \delta_i \frac{T^4}{\Tbh^2} \left[ 1 - \frac{\epsilon}{\delta_i} \left( \frac{\Tbh}{T} \right)^4 \right]\,,
    \label{Eq:evolution}
\eeq

\noi
where $\epsilon \equiv \frac{27}{4}\left(\frac{\pi}{480}\right)
g_*(T_{\rm BH})$ characterizes the evaporation rate, while
$\delta_i$ ($i={\rm lf}$ or ${\rm hf}$) denotes the absorption
coefficient in the low- and high-frequency regimes, respectively.
In the high-frequency limit, $\delta_i$ is independent of the spin of the
absorbed particle, whereas in the low-frequency regime it depends on
the particle spin $s$:

\beq
\delta_{\rm hf}=\frac{9\pi}{640}g_*\,,
\eeq

\beq
\delta_{\rm lf}\sim
\left\{
\begin{aligned}
&\left(\frac{\pi}{120}g_*\right)\,,~~~~~~~s=0\,,\\
&\left(\frac{\pi}{960}g_*\right),~~~~~~~~~~s=1/2\,,\\
&\left(\frac{\rm 1}{\rm 5760 \pi}\frac{\omega^2}{\Tbh^2}g_*\right),~~~~~s=1\,,
\label{Eq:deltalf}
\end{aligned}\right.
\eeq.

\noi
with $g_*$ the effective degrees of freedom in the thermal bath.
The factor $\frac{27}{4}$  in $\epsilon$ accounts for the greybody factor \cite{Arbey:2019mbc, Cheek:2021odj, Baldes:2020nuv}.

In Eqs.~(\ref{Eq:evolution}) and (\ref{Eq:deltalf}),  the Hawking temperature $\Tbh$ is defined by 

\beq
\Tbh=\frac{M_P^2}{\Mbh}\,,
\eeq

\noi
where $M_P = 1/\sqrt{8\pi G} \simeq 2.4 \times 10^{18} \, \rm{GeV}$ is the reduced Planck mass \cite{Hawking:1975vcx}.

The evolution of a primordial black hole then results from the competition between absorption from the surrounding thermal bath and Hawking evaporation. The evaporation contribution is encoded in the second term on the right-hand side of Eq.~(\ref{Eq:evolution}) and, when considered separately, is given by

\beq
\frac{d \Mbh}{dt} = -\epsilon \frac{M_P^4}{\Mbh^2}=-\epsilon \Tbh^2\,.
\label{Eq:evaporation}
\eeq

\noi
To solve Eq.~(\ref{Eq:evolution}), one must specify the initial PBH mass at the time of formation. Assuming that PBHs form with a fraction $\gamma$ of the horizon mass during the radiation-dominated era, the initial mass is given by
\bea
\Min&&=\gamma \rho_{\rm R}(t_{\rm in})\frac{4}{3}\pi\frac{1}{H_{\rm in}^3}=4\pi\gamma\frac{M^2_P}{H_{\rm in}}
\nonumber
\\
&&
\simeq 1.3~\gamma ~{\rm g}\left(\frac{10^{14}~\rm GeV}{\Hin}\right)
\,,
\label{Eq:min}
\eea

\noi
where $\gamma$ is the collapse efficiency parameter\footnote{It is worth noting that more accurate methods exist for estimating the formation mass of primordial black holes (PBHs), which take into account both the nature of the cosmological background and the specific profile of scalar perturbations~\cite{Musco:2012au, Musco:2008hv, Hawke:2002rf, Niemeyer:1997mt, Escriva:2021pmf, Escriva:2019nsa, Escriva:2020tak, Escriva:2021aeh}. However, a fully analytical understanding of the values of $\gamma$, particularly in the context of PBH formation during the post-inflationary epoch is still lacking.}.
Here, $H_{\rm in}$ denotes the Hubble parameter at the time of PBH formation, which is related to the background radiation temperature in a radiation dominated Universe:

\beq
H_{\rm in} = \sqrt{\frac{\rho_R (t_{\rm in})}{3M_P^2}} = 
\sqrt{\frac{\alpha}{3}}\,\frac{T_{\rm in}^2}{M_P}\,,
\label{Eq:hin}
\eeq

\noi
with $\alpha=g_* (T)\frac{\pi^2}{30}$.
$T_{\rm in}$ is the radiation temperature at the time of formation.

\subsection{Absorption during reheating}

It was shown in Ref.~\cite{Haque:2026vvp} that, in a radiation-dominated Universe, there exists a critical value of $\gamma$,

\beq
\gamma=\gamc =\frac{\alpha}{6 \pi \delta_i} \simeq 0.395\,,
\label{Eq:gammacr}
\eeq

\noi
above which PBHs undergo a phase of runaway absorption and grow without bound. 

Before proceeding, we emphasize that we do not address the possible limitations of the runaway regime in this work. Instead, we focus on the existence and determination of the critical collapse efficiency, $\gamma_c$ in case of PBH formed during reheating, while acknowledging that, above this threshold, the PBH enters a runaway growth regime \cite{Barrau:2022bfg}. We also stress that the critical values derived throughout this work rely on the assumption of instantaneous PBH formation. Relaxing this assumption and allowing for a finite collapse time could quantitatively modify the critical collapse efficiencies obtained here.

The result of Eq.~\eqref{Eq:gammacr}, however, crucially relies on the assumption that PBHs form during a radiation-dominated era. In realistic cosmological scenarios, PBH formation may instead occur during reheating \cite{Mambrini:2021gpp, Garcia:2020eof, Kaneta:2019zgw,Haque:2020zco}, when the expansion of the Universe is driven by the oscillating inflaton field $\phi$ with energy density $\rho_\phi$, rather than by the radiation bath. As a consequence, the evolution of the Hubble rate is modified. The inflaton energy density during reheating scales as

\beq
\rho_\phi \propto a^{-3(1+w_\phi)}\,,
\eeq 
\noi
where
\beq
w_\phi=\frac{k-2}{k+2}\,,
\eeq
\noi
for a potential $V(\phi)\sim \phi^k$.

Assuming that reheating produces a thermal bath\footnote{ We assume that the decay products of the inflaton instantaneously thermalize.}, with the temperature depending on the scale factor $a$ as, 

\beq
T=\Trh\left(\frac{\arh}{a}\right)^\zeta\,,
\eeq

\noi
where $\Trh$ is the reheating temperature defined by 
\beq
\rho_\phi(\arh)=\rho_R(\arh)=\alpha \Trh^4\,.
\label{Eq:rhorh}
\eeq
\noi
Note that for reheating to be possible, and thus for condition (\ref{Eq:rhorh}) to be reached, the density of the inflaton, $\rho_\phi$, must be diluted more efficiently than the radiation density,
$\rho_R=\alpha T^4$, which translates to
\beq 
4 \zeta < 3 + 3w_\phi \,. \label{Eq:conditionrh} 
\eeq
\noi

While this condition might look arbitrary in our parametrization, we emphasize that a Lagrangian model of reheating would provide $\zeta$ as a function of the equation of state $w_{\phi}$, thereby constraining the parameters of the model \cite{Garcia:2023obw,Garcia:2020wiy,Garcia:2020eof}. With this in mind, the benchmark points used later correspond to specific reheating models.

Within the inflaton-dominated approximation during reheating,
the Hubble rate evolves as
\beq
H=\Hrh\left(\frac{\arh}{a}\right)^{\frac{3}{2}(1+w_\phi)}\,.
\label{Eq:Ha}
\eeq
\noi
It is convenient to normalize the PBH mass to its formation value and define
$R=\frac{\Mbh}{\Min}$. We can then rewrite Eq.~(\ref{Eq:evolution}) as

\beq
\frac{dR}{R^2} = \delta_i\frac{da}{a}\frac{\Min}{H}\frac{T^4}{M_P^4}\left[ 1 - \frac{\epsilon}{\delta_i} \left( \frac{\Tbh}{T} \right)^4 \right]\,,
\eeq

\noi
or by defining $x=\frac{a}{\arh}$ and focusing initially on the absorption phase\footnote{ This is justified as the condition for a dominant absorption prior to evaporation is unchanged during reheating \cite{Haque:2026vvp}}, 

\beq
\frac{dR}{R^2}=\delta_i\frac{\Min\Trh^4}{M_P^4\Hrh}x^{\frac12+\frac32 w_\phi-4 \zeta} dx \,.
\label{Eq:drovr2}
\eeq

Integrating this last equation between $\xin$ and $x$ yields a solution of the form, for $3+3w_\phi-8\zeta \neq 0$, 
\beq
\frac{1}{R}=1-\frac{2 \delta_i}{3+3w_\phi-8 \zeta}\sqrt{\frac{3}{\alpha}}\frac{\Min \Trh^2}{M_P^3}\left[x^{\frac{3+3w_\phi-8\zeta}{2}}-\xin^{\frac{3+3w_\phi-8 \zeta}{2}}\right],
\label{Eq:sol1}
\eeq

\noi
and
\beq
\frac{1}{R}= 1-\delta_i \sqrt{\frac{3}{\alpha}}\frac{\Min \Trh^2}{M_P^3}\left[
\ln(x)-\ln(\xin)\right] \,,
\label{Eq:sollog}
\eeq
\noi 
for $3+3w_\phi-8\zeta = 0$. 
Eq.~(\ref{Eq:sol1}) already shows that runaway absorption occurs whenever the 
denominator reaches zero before the end of reheating. Whether this happens depends on the PBH formation time and on the cosmological background through the combination
\beq
D\equiv 3+3w_\phi-8\zeta\,.
\eeq

For $D<0$, the absorption integral is dominated by its lower limit, and the mass growth is therefore UV dominated, i.e. controlled mainly by the earliest times after PBH formation. Conversely, for $D>0$, the integral is dominated by its upper limit, and the growth is IR dominated, being controlled mainly by times close to the end of reheating. The limiting case $D=0$ gives the logarithmic evolution of Eq.~(\ref{Eq:sollog}).

Since PBHs are assumed to form {\it during} reheating, one must require
$\xin<1$, where 
\beq
\xin=\frac{\ain}{\arh}=\left(\frac{\rhorh}{\rhoin}\right)^\frac{1}{3+3w_\phi}=\left(\frac{\alpha \Trh^4\Min^2}{48 \pi^2\gamma^2M_P^6}\right)^\frac{1}{3+3w_\phi}\,,
\label{Eq:xin}
\eeq

\noi
the condition $\xin<1$ becomes
\beq
\Min<\frac{M_P^3}{\Trh^2}\sqrt{\frac{3}{\alpha}}4 \pi \gamma\simeq 1.8\times 10^{11}\left(\frac{10^{10}~\rm GeV}{\Trh}\right)^2\frac{\gamma}{0.2}~\rm g\,.
\label{Eq:Minmax}
\eeq

\noi
For any given reheating temperature $\Trh$, there is therefore a finite 
range of initial PBH masses satisfying $\xin<1$. In particular, even 
for very high reheating temperatures, at most $\Trh\sim 10^{15}~{\rm GeV}$, corresponding to an inflationary energy density of order $\rho_\phi\sim 10^{60}~{\rm GeV}^4$, the upper bound remains of order
\beq
\Min \lesssim {\cal O}(10)~{\rm g}\left(\frac{\gamma}{0.2}\right)\,,
\eeq

\noi
so that PBHs with sufficiently small initial masses can consistently 
form during reheating. In what follows, we restrict ourselves to this 
regime, for which the relevant evolution takes place at $x<1$.

\subsection{Breakdown of the universality of the critical collapse efficiency $\gamc$}

\subsubsection{Runaway during reheating}

The existence of a reheating phase immediately raises a fundamental
question: does the universal critical collapse efficiency derived in a
radiation-dominated Universe, $\gamc\simeq0.395$, survive in a realistic
cosmological history, or is it modified by the mass absorbed before the
onset of radiation domination?
As 
we now demonstrate, reheating breaks the  universality observed in radiation domination. During radiation 
domination, the critical collapse efficiency $\gamc$ is universal and does 
not depend on the initial PBH mass $\Min$, as shown in \cite{Haque:2026vvp}. This universality reflects the scaling properties of the
radiation-dominated background, for which no additional cosmological
scale enters the critical condition. 

By contrast, during reheating the amount of mass absorbed depends on the
PBH formation time or, for a fixed collapse efficiency and reheating
background, equivalently on its initial mass, since the ratio
$\rho_{\rm rad}/\rho_{\rm tot}$ evolves with time. Consequently, although the universal critical threshold of the radiation-dominated era itself remains unchanged, the corresponding critical collapse efficiency at PBH formation no longer does. Instead, it depends on the absorption history during reheating, encoded in the parameter $\xin=\ain/\arh$.
 We first illustrate this breakdown of universality by considering the most extreme situation, 
namely that the PBH already enters the runaway regime during reheating itself.

Using Eq.~(\ref{Eq:xin}),
\beq
    \frac{\Min \Trh^2}{M_P^3}
    =
    4\pi\gamma\sqrt{\frac{3}{\alpha}}\,
    \xin^\frac{3 + 3w_\phi}{2}\,,
    \label{Eq:MiT2MP3}
\eeq

\noi which transforms Eq.(\ref{Eq:drovr2}) into 

\bea
\frac{dR}{R^2}
&=&
\delta_i
\frac{\Min\Trh^2}{M_P^3}
\sqrt{\frac{3}{\alpha}}\,
x^{\frac12+\frac32 w_\phi-4 \zeta} dx
\nonumber\\
&=&
2\,\frac{\gamma}{\gamc}\,
\xin^{\frac{3+3w_\phi}{2}}
x^{\frac12+\frac32 w_\phi-4 \zeta} dx\,,
\label{Eq:drovr2bis}
\eea

\noi
It follows that

\beq
    \frac{1}{\Rrh}
    =
    1
    -
    \frac{4}{3+3w_\phi-8\zeta}
    \frac{\gamma}{\gamc}
    \left[
    \xin^\frac{3+3w_\phi}{2}
    -
    \xin^\frac{6+6w_\phi-8\zeta}{2}
    \right]\,.
    \label{Eq:Rrh}
\eeq
Eq.~(\ref{Eq:Rrh}) already reveals the possibility of runaway
absorption: when $1/\Rrh$ vanishes, the perturbative solution formally
diverges before the end of reheating.

The origin of the $\xin$ dependence is simple. In the radiation-dominated
analysis of Ref.~\cite{Haque:2026vvp}, PBHs are assumed to form directly
during radiation domination, so that no additional cosmological scale
enters the problem. During reheating, however, the reheating transition introduces an
additional physical scale, characterized for instance by $H_{\rm RH}$
or $T_{\rm RH}$. The PBH formation time relative to this transition is
conveniently characterized by the dimensionless ratio
$\xin=a_{\rm in}/a_{\rm RH}$. Since the initial PBH mass depends on the
Hubble rate at formation,
\beq
M_{\rm in}\propto\gamma H_{\rm in}^{-1},
\eeq
and
\beq
H_{\rm in}
=
H_{\rm RH}\,
\xin^{-\frac32(1+w_\phi)},
\eeq
one immediately obtains
\beq
M_{\rm in}
\propto
\gamma\,
\xin^{\frac32(1+w_\phi)}.
\eeq
The PBH mass at formation, and consequently its subsequent absorption
history, therefore retain an explicit memory of the formation time during
reheating.

Equation~(\ref{Eq:Rrh}) immediately shows that the perturbative solution
formally diverges when the denominator of $\Rrh$ vanishes. This signals
the onset of a non-perturbative absorption regime {\it during} reheating,
corresponding to runaway PBH growth before the onset of radiation
domination. For $3+3w_\phi-8\zeta\neq0$, imposing the condition
\beq
\frac{1}{\Rrh}=0\,,
\eeq
one obtains the critical collapse efficiency required for runaway growth
during reheating,
\beq
\gamma_{c}^{\rm RH}
=
\gamc\,
\frac{
(3+3w_\phi-8\zeta)\,
\xin^{-\frac{3+3w_\phi}{2}}
}
{
4\left[
1-\xin^{\frac{3+3w_\phi-8\zeta}{2}}
\right]
}\,.
\label{Eq:gammarunreh}
\eeq

\noi
For $3+3w_\phi-8\zeta=0$, the power-law solution is replaced by the
logarithmic one, which gives
\beq
\frac{1}{R_{\rm RH}}
=
1+
2\,\frac{\gamma}{\gamc}\,
\xin^{\frac{3}{2}(1+w_\phi)}
\ln(\xin)\,,
\eeq
and the corresponding critical collapse efficiency for runaway
during reheating is
\beq
\gamma_{c,{\rm log}}^{\rm RH}
=
-\frac{\gamc}{2}\,
\frac{\xin^{-\frac{3}{2}(1+w_\phi)}}{\ln(\xin)}\,.
\label{Eq:gammarunreh_log}
\eeq

\noi
For a collapse efficiency larger than the corresponding critical value,
the perturbative solution breaks down and the PBH enters the runaway
regime before the end of reheating.

In the limit in which the PBH forms much earlier than the end of
reheating, $\xin\ll1$, this expression admits simple asymptotic forms.
For $3+3w_\phi-8\zeta>0$, the second term in the denominator of
Eq.~(\ref{Eq:gammarunreh}) becomes negligible, yielding
\beq
\gamma_{c}^{\rm RH}
\simeq
\frac{\gamc}{4}
(3+3w_\phi-8\zeta)\,
\xin^{-\frac{3+3w_\phi}{2}}\,.
\label{Eq:gammarunreh_xinsmall_pos}
\eeq

\noi
For $3+3w_\phi-8\zeta<0$, the denominator is instead dominated by
$\xin^{(3+3w_\phi-8\zeta)/2}$. One then obtains
\beq
\gamma_{c}^{\rm RH}
\simeq
\frac{\gamc}{4}
\left|3+3w_\phi-8\zeta\right|\,
\xin^{\,4\zeta-3-3w_\phi}\,.
\label{Eq:gammarunreh_xinsmall_neg}
\eeq

\begin{figure}
    \centering
    \includegraphics[width=\linewidth]{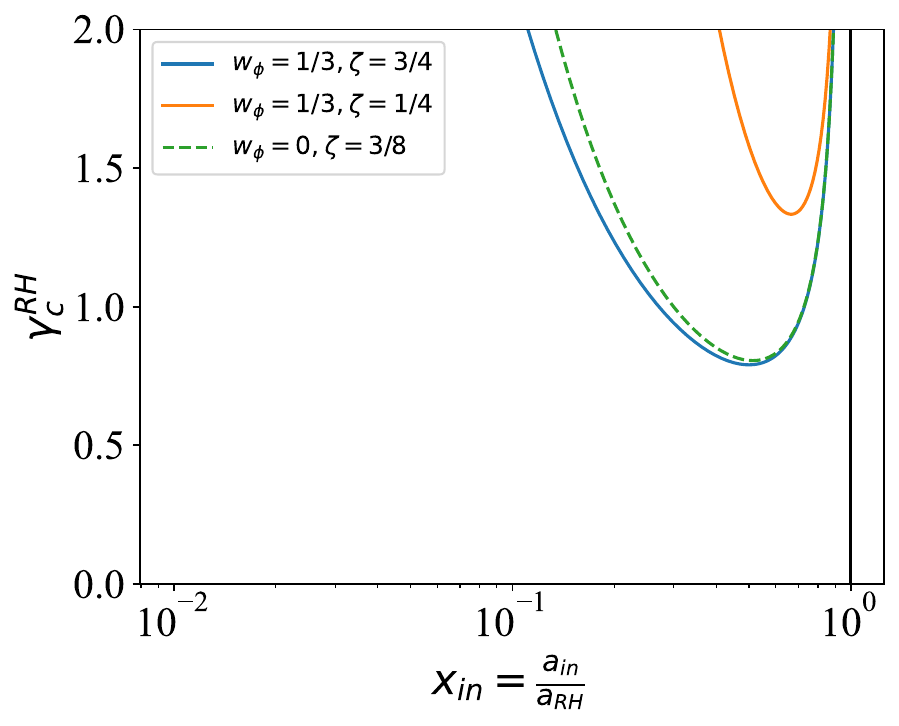}
    \caption{\justifying Critical collapse efficiency $\gamma_c^{\rm RH}$ required for runaway
    absorption to occur before the end of reheating, obtained from Eq.~(\ref{Eq:gammarunreh}),
$3+3w_\phi-8\zeta\neq0$, as a function of the PBH formation time
    $x_{\rm in}=a_{\rm in}/a_{\rm RH}$. The solid blue and orange curves
    correspond respectively to representative UV-dominated
    ($w_\phi=1/3$, $\zeta=3/4$) and IR-dominated
    ($w_\phi=1/3$, $\zeta=1/4$) reheating histories. The dashed green curve represents the logarithmic case
$3+3w_\phi-8\zeta=0$, with $w_\phi=0$ and $\zeta=3/8$,
obtained from Eq.~(\ref{Eq:gammarunreh_log}).
    }
    \label{fig:duringRH}
\end{figure}

We show in Fig.~\ref{fig:duringRH} the critical collapse efficiency
$\gamma_c^{\rm RH}$ required for a PBH to enter the runaway regime
before the end of reheating. The blue curve corresponds to the UV-dominated regime,
with $w_\phi=1/3$ and $\zeta=3/4$, as obtained for an
inflaton oscillating in a quartic potential and reheating either
through perturbative decays into fermions,
$\phi\rightarrow\bar f f$, or through inflaton annihilations into
bosons, $\phi\phi\rightarrow bb$. The orange curve shows the IR-dominated regime, with
$w_\phi=1/3$ and $\zeta=1/4$, corresponding to perturbative
inflaton decays into bosons, $\phi\rightarrow bb$, for a quartic
inflaton potential. Finally, the dashed
green curve represents the logarithmic case,
$3+3w_\phi-8\zeta=0$, obtained for $w_\phi=0$ and $\zeta=3/8$, as
expected for perturbative reheating after inflation in a quadratic
potential regardless of the coupling achieving the reheating\cite{Garcia:2023obw,Garcia:2020wiy,Garcia:2020eof}.

In all three cases, $\gamma_c^{\rm RH}$ increases both for
$x_{\rm in}\ll1$ and for $x_{\rm in}\rightarrow1$, implying that runaway
during reheating is most easily achieved for PBHs formed at an
intermediate stage of reheating. The increase at small $x_{\rm in}$
reflects the smaller initial mass of earlier-forming PBHs, whereas the
divergence for $x_{\rm in}\rightarrow1$ simply indicates that too little
time remains for runaway absorption to develop before the end of
reheating. This divergence is therefore an artifact of the present criterion rather
than a physical effect. In the next section, we instead determine the
critical collapse efficiency $\gamma_c^{\rm RH}$ required for the PBH to
reach the runaway regime at the onset of radiation domination. As
expected, $\gamma_c^{\rm RH}$ smoothly approaches $\gamc$ in the limit
$x_{\rm in}\rightarrow1$.

For sufficiently early formation, $\xin\ll1$, the critical value
$\gamma_c^{\rm RH}$ is pushed to large values. PBHs formed earlier during
reheating correspond to larger $H_{\rm in}$ and hence to a smaller
initial horizon mass,
$\Min\propto\gamma H_{\rm in}^{-1}$. The resulting mass growth through
absorption is therefore suppressed, so that a larger collapse efficiency
$\gamma$ is required for the PBH to enter the runaway regime before the
end of reheating.

The above condition corresponds to the most extreme outcome of PBH growth during reheating. 
Remarkably, however, runaway absorption during reheating is not the relevant
criterion for the subsequent evolution. Even if the PBH remains in the
perturbative regime throughout reheating, the accumulated mass modifies its
effective collapse efficiency at the onset of radiation domination.


\subsubsection{Runaway at the onset of radiation domination}

\subsubsection*{The case $3+3w_\phi-8\zeta\neq0$}

Naively, one might expect that the onset of a non-perturbative 
absorption regime {\it during} reheating directly determines the critical 
collapse efficiency. However, this is not the relevant criterion. Even 
if the PBH remains in the perturbative regime throughout reheating, the mass absorbed during this epoch increases the PBH mass at the onset of 
radiation domination. Since the runaway threshold during 
radiation domination is highly sensitive to the PBH mass, this 
additional mass can already be sufficient to trigger runaway growth 
after reheating. We now quantify this effect.

For that, it is convenient to introduce an effective collapse 
efficiency, $\gameff$, defined as {\it the value of the collapse parameter 
that would produce the same PBH mass if formation occurred directly at the onset of radiation domination}. Using Eq.~(\ref{Eq:min}), one writes

\beq
\Mrh
=
4\pi \gameff
\frac{M_P^2}{\Hrh}
\,\Leftrightarrow\,
\gameff
=
\frac{1}{4\pi}
\sqrt{\frac{\alpha}{3}}
\frac{\Mrh \Trh^2}{M_P^3}\,.
\eeq

\noi
Using $\Mrh=\Rrh\Min$ together with Eq.~(\ref{Eq:xin}), one finds

\beq
\gameff
=
\gamma\,\Rrh\,
\xin^\frac{3+3w_\phi}{2}\,.
\label{Eq:gammaeff}
\eeq

The onset of runaway growth during radiation domination is therefore determined by the universal condition

\beq
\gameff=\gamc\,,
\eeq

\noi
which implies

\beq
\gamma\,\Rrh\,
\xin^\frac{3+3w_\phi}{2}
=
\gamc\,.
\label{Eq:gammacritreh}
\eeq

\noi
\noi
Using the expression (\ref{Eq:Rrh}) for $\Rrh$, the condition 
$\gameff \geq \gamc$ can be rewritten as

\beq
\gamma \geq\frac{\alpha}{6 \pi \delta_i}
\xin^{-\frac{3+3w_\phi}{2}}
-
\frac{4 \gamma}{3+3w_\phi - 8\zeta}
\left(
1 - \xin^{\frac{3+3w_\phi-8\zeta}{2}}
\right)\,.
\eeq

\noi
Solving for $\gamma$, we obtain the critical value
\beq
\gamma  \geq\gamma_c^{\rm RD}=
\gamc
\frac{
(3+3w_\phi-8\zeta)\,
\xin^{-\frac{3+3w_\phi}{2}}
}
{
7+3w_\phi-8\zeta
-
4\xin^{\frac{3+3w_\phi-8\zeta}{2}}
}\,.
\label{Eq:gammacritreh2}
\eeq

\noi
Note that this criterion is very different from the one obtained by requiring runaway growth {\it during} reheating, given in Eq.~(\ref{Eq:gammarunreh}).
This is expected, since 
Eq.~(\ref{Eq:gammarunreh}) requires the perturbative solution to {\it diverge before} the end of reheating. By contrast, 
Eq.~(\ref{Eq:gammacritreh2}) only requires the PBH to reach the universal 
radiation-era threshold at the onset of radiation domination, naturally recovering the result of \cite{Haque:2026vvp},

\beq
\gamma
\longrightarrow
\gamc
=
\frac{\alpha}{6\pi\delta_i}=0.395\,.
\eeq
\noi
when $\xin \rightarrow 1$.

In the limit $\xin\ll1$, the critical condition admits simple asymptotic forms. For
$3+3w_\phi-8\zeta>0$, the last term in the denominator of
Eq.~(\ref{Eq:gammacritreh2}) becomes negligible, and one obtains
\beq
\gamma
 \gtrsim
\gamc\,
\frac{
3+3w_\phi-8\zeta
}
{
7+3w_\phi-8\zeta
}
\xin^{-\frac{3+3w_\phi}{2}}\,.
\label{Eq:gammacritreh2_xinsmall_pos}
\eeq
Comparing Eq.~(\ref{Eq:gammacritreh2_xinsmall_pos}) with
Eq.~(\ref{Eq:gammarunreh_xinsmall_pos}), one finds that the two critical
collapse efficiencies exhibit the same asymptotic dependence on $\xin$,
but differ by a constant prefactor,
\beq
\frac{\gamma_c^{\rm RH}}{\gamma_c^{\rm RD}}
=
\frac{7+3w_\phi-8\zeta}{4}
=
1+\frac{3+3w_\phi-8\zeta}{4}
>1\,.
\eeq
The common scaling with $\xin$ reflects the suppression of the initial PBH
mass for very early formation. The larger prefactor for
$\gamma_c^{\rm RH}$ follows from the more stringent requirement that the
PBH enter the non-perturbative regime before reheating is completed,
whereas $\gamma_c^{\rm RD}$ only requires the accumulated reheating growth
to bring the PBH to the universal radiation-era threshold at the onset of
radiation domination.

For $3+3w_\phi-8\zeta<0$, the denominator is instead dominated by the last term,
and the critical condition becomes

\beq
\gamma_c^{\rm RD}
\gtrsim
\frac{\gamc}{4}
\left|3+3w_\phi-8\zeta\right|
\xin^{\,4\zeta-3-3w_\phi}\,,
\label{Eq:gammacritreh2_xinsmall_neg}
\eeq

Remarkably, this asymptotic expression is identical to the corresponding
runaway criterion during reheating,
Eq.~(\ref{Eq:gammarunreh_xinsmall_neg}).
This exact agreement is not accidental. It reflects the fact that in the regime $3+3w_\phi-8\zeta<0$, the absorption integral is UV dominated,
so that the PBH acquires nearly all of the mass gained during reheating
immediately after its formation.
As a result, requiring the PBH to enter the runaway regime during
reheating or only at the onset of radiation domination becomes
asymptotically equivalent.

\subsubsection*{The case $3+3w_\phi-8\zeta = 0$}

The previous discussion assumed
$3+3w_\phi-8\zeta\neq0$.
We now turn to the degenerate case
\beq
3+3w_\phi-8\zeta=0\,,
\eeq
for which the power-law solution is replaced by a logarithmic behaviour.
Evaluating Eq.~(\ref{Eq:sollog}) at $x=1$ and using Eq.~(\ref{Eq:MiT2MP3}) gives

\beq
\frac{1}{\Rrh}
=
1
+
\frac{12 \pi \gamma \delta_i}{\alpha}
\xin^{\frac{3+3w_\phi}{2}}
\ln(\xin)\,,
\eeq
\noi
thus implying 
\beq
\gamma_{c}^{\rm RH} = -\frac{\gamma_{c}}{2}\frac{x^{-\frac{3(1+w_{\phi})}{2}}}{\mathrm{ln}(x)}
\label{Eq:gammac log}
\eeq
Using the definition of the effective collapse efficiency, Eq.~(\ref{Eq:gammaeff}), the critical condition becomes

\beq
\gamma
\geq
\frac{\alpha}{6 \pi \delta_i}
\xin^{-\frac{3+3w_\phi}{2}}
+
2\gamma \ln(\xin)\,.
\eeq

Solving for $\gamma$ yields

\beq
\gamma
\geq
\frac{\alpha}{6 \pi \delta_i}
\frac{\xin^{-\frac{3+3w_\phi}{2}}}
{1-2\ln(\xin)}
=
\gamc\,
\frac{\xin^{-\frac{3+3w_\phi}{2}}}
{1-2\ln(\xin)}\,.
\label{Eq:gammacritlog}
\eeq

Equation~(\ref{Eq:gammacritlog}) appears to develop a shallow minimum
slightly below the universal threshold $\gamc$.
This feature, however, is an artifact of approximating the expansion
history by two separate phases, using only the dominant energy component
to determine the Hubble rate in each regime.
When the exact background evolution,
\beq
H^2=\frac{\rho_\phi+\rho_R}{3M_P^2}\,,
\eeq
is used instead, this spurious minimum disappears.
The critical collapse efficiency remains bounded from below by the
universal radiation-dominated value,
\beq
\gamma_c^{\rm RD}\ge\gamc,
\eeq
with equality recovered in the limit
$\xin\rightarrow1$.

\begin{figure}
    \centering    \includegraphics[width=\linewidth]{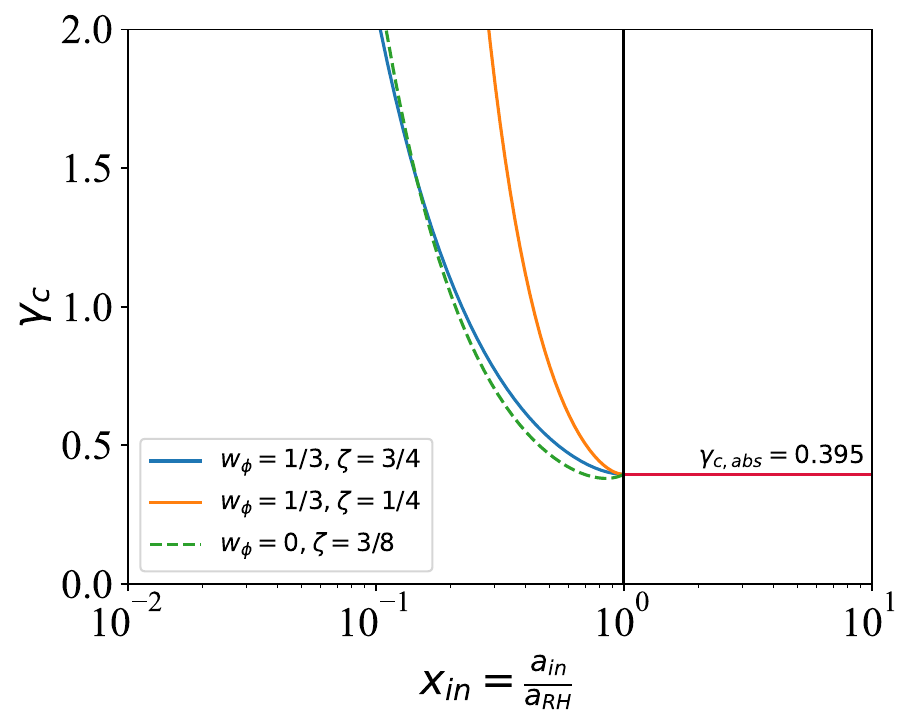}
    \caption{ \justifying
Critical collapse efficiency $\gamma_c^{\rm RD}$ required for a PBH formed
during reheating to reach the runaway regime at the onset of radiation
domination, as a function of the formation time
$\xin=a_{\rm in}/a_{\rm RH}$. The benchmark reheating histories are the
same as in Fig.~\ref{fig:duringRH}. The vertical line marks the onset of
radiation domination, while the horizontal line corresponds to the
universal radiation-dominated threshold
$\gamma_c^{\rm RD}=\gamc=0.395$ from Ref.~\cite{Haque:2026vvp}.
}
    \label{fig:RHandRD}
\end{figure}

In Fig.~\ref{fig:RHandRD}, we show the critical collapse efficiency
required for a PBH formed during reheating to reach the runaway regime at
the onset of radiation domination for the three benchmark reheating
histories discussed above.

All solutions smoothly approach the universal radiation-dominated
threshold,
\beq
\gamma_c^{\rm RD}=\gamc,
\eeq
in the limit $\xin\rightarrow1$, providing a non-trivial consistency
check of the analytical treatment.

For $3+3w_\phi-8\zeta>0$, the critical collapse efficiency required to
trigger runaway during reheating remains larger than
$\gamma_c^{\rm RD}$ by a constant prefactor.
By contrast, for $3+3w_\phi-8\zeta<0$, the two criteria become
asymptotically identical because the reheating-induced mass growth is
dominated by the earliest stages following PBH formation.

These results highlight the physical difference between the two
criteria.
Equation~(\ref{Eq:gammarunreh}) requires the perturbative solution to
break down before reheating ends, whereas
Eq.~(\ref{Eq:gammacritreh2}) only requires the perturbative mass growth
during reheating to bring the PBH above the universal radiation-era
runaway threshold.
Consequently, the relevant quantity is not whether the perturbative solution reaches
its runaway singularity during reheating, but whether the accumulated perturbative growth is sufficient to trigger runaway once radiation domination begins.
This constitutes one of the main results of this work.
\section{Generalization to multiple mass-growth processes and successive cosmological eras}\label{Sec:generalization}

The results derived in the previous section are not restricted to the
specific case of reheating followed by radiation domination.
They can be cast into a general composition framework
applicable to cosmological histories involving successive epochs and
multiple PBH mass-growth mechanisms. In general, several independent mass-growth mechanisms may contribute
simultaneously to the evolution of a primordial black hole within a
given cosmological era. Moreover, the mass accumulated during one era
modifies the initial condition for the subsequent evolution after a
change in the cosmological background. We first derive a general addition rule describing how
independent mass-growth mechanisms combine within a given cosmological
era, and then establish the corresponding matching prescription between
successive cosmological eras. We subsequently apply this framework to
the reheating epoch, where radiation absorption and inflaton accretion
act simultaneously.
\subsection{Multiple mass-growth processes}

During a cosmological era dominated by a component with
constant equation-of-state parameter $w$, the Hubble rate evolves as
\beq
H\propto a^{-\frac{3}{2}(1+w)}\,.
\eeq
Equation~(\ref{Eq:evolution}) is a particular realization of a broader
class of PBH mass-growth processes. Neglecting the evaporation term, the
contribution of a generic process $i$ can be written as
\beq
\frac{d\Mbh}{dt}
=
\xi_i\,\rho_i\,\frac{\Mbh^2}{\Mp^4}\,,
\label{Eq:massgrowthera}
\eeq
where $\rho_i$ denotes the energy density of the species driving the
mass-growth process, $\xi_i$ is its associated efficiency and
$\Mbh^{2}/\Mp^{4}$ sets the characteristic geometrical area
scaling of the black hole. For the
radiation-absorption process studied above,
$\rho_i=\rho_R$ and
$\xi_i=30\,\delta_i/(\pi^2g_*)$.

We now allow several independent mass-growth processes to act
simultaneously during the same cosmological era. To derive the
corresponding evolution equation, it is convenient to introduce
\beq
x=\frac{a}{a_{\rm tr}}\,,
\eeq
where $a_{\rm tr}$ denotes the transition scale factor at the end of the
era. Since $\dot{x}=Hx$, one has
\beq
\frac{d\Mbh}{dt}
=
Hx\frac{d\Mbh}{dx}
=
M_{\rm in}Hx\frac{dR}{dx}\,.
\eeq
Using
\beq
\frac{d}{dx}\left(\frac1R\right)
=
-\frac1{R^2}\frac{dR}{dx}\,,
\eeq
the contribution of process $i$ can be written as
\beq
\frac{d}{dx}\left(\frac1R\right)_i
=
-
\frac{\xi_i\rho_i}{\Mp^4}
\frac{M_{\rm in}}{Hx}\,.
\label{Eq:d1Rdxgeneric}
\eeq

\noi
We recall that for a PBH formed at $x=x_{\rm in}$, its initial mass is parametrized as
Eq.~(\ref{Eq:min}), $M_{\rm in}=4\pi\gamma\Mp^2/H_{\rm in}$,
where $\gamma$ is fixed by the
formation process and {\it should not be confused} with the mass-growth
efficiency $\xi_i$ associated with the subsequent process $i$. With
\beq
\frac{H}{H_{\rm in}}
=
\left(
\frac{x}{x_{\rm in}}
\right)^{-\frac{3}{2}(1+w)}\,,
\eeq
and writing the energy density of species $i$ as
\beq
\rho_i
=
\rho_{i,{\rm tr}}\,
x^{-\chi_i}\,,
\label{Eq:rhophievolution}
\eeq
one obtains
\beq
\frac{d}{dx}
\left(
\frac1R
\right)_i
=
-
\gamma\,
{\cal A}_i\,
x^{-\chi_i+\frac32(1+w)-1}\,,
\eeq
with
\beq
{\cal A}_i
=
4\pi\,
\xi_i\,
\frac{\rho_{i,\rm tr}}
{\Mp^2H_{\rm in}^2}\,
x_{\rm in}^{-\frac32(1+w)}\,.
\label{Eq:calA}
\eeq

\noi
The coefficient ${\cal A}_i$ contains all the normalization factors,
which depend on the initial conditions and on the
mass-growth process $i$.

When several independent mass-growth mechanisms are simultaneously
present, the total PBH mass-growth rate is assumed to be the sum of the
individual contributions,
\beq
\frac{d\Mbh}{dt}
=
\sum_i
\left(
\frac{d\Mbh}{dt}
\right)_i
=
\sum_i
\xi_i\rho_i
\frac{\Mbh^2}{\Mp^4}\,.
\eeq
Since Eq.~(\ref{Eq:d1Rdxgeneric}) is linear in the source term
$\xi_i\rho_i$, the total evolution equation immediately becomes
\beq
\frac{d}{dx}
\left(
\frac1R
\right)_{\rm tot}
=
\sum_i
\frac{d}{dx}
\left(
\frac1R
\right)_i\,.
\eeq

Integrating from PBH formation, $x=x_{\rm in}$, to the end of the era,
$x=1$, gives
\beq
\frac1{R_{\rm tr}}
=
1
-
\gamma
\sum_i
{\cal I}_i\,,
\eeq
where
\beq
{\cal I}_i
\equiv
{\cal A}_i
\int_{x_{\rm in}}^{1}
x^{-\chi_i+\frac32(1+w)-1}\,
dx\,,
\label{Eq:calI}
\eeq

\noi
is the integrated contribution of process $i$.

When only process $i$ is present,
\beq
\frac1{R_{\rm tr}}
=
1-\gamma{\cal I}_i\,.
\eeq
The critical collapse efficiency associated with this process
acting alone is
obtained by requiring $1/R_{\rm tr}=0$, yielding
\beq
\frac1{\gamma_{c,i}}
=
{\cal I}_i\,.
\eeq
One therefore obtains
\beq
\frac1{R_{\rm tr}}
=
1
-
\gamma
\sum_i
\frac1{\gamma_{c,i}}\,.
\eeq
The combined critical collapse efficiency during the considered era is
thus defined by
\beq
\frac1{\gamma_{c,{\rm era}}}
=
\sum_i
\frac1{\gamma_{c,i}}\,,
\eeq
so that
\beq
\frac1{R_{\rm tr}}
=
1
-
\frac{\gamma}{\gamma_{c,{\rm era}}}\,.
\label{Eq:Rgenericera}
\eeq

Thus, different mass-growth mechanisms can be studied independently and
combined through the simple addition rule
\beq
\boxed{
\frac1{\gamma_{c,{\rm era}}}
=
\sum_i
\frac1{\gamma_{c,i}}
}
\,,
\label{Eq:addition}
\eeq
which constitutes the addition theorem.
The structure of this theorem is independent of the detailed
cosmological evolution. All the dependence on the background history is
encoded in the individual critical efficiencies $\gamma_{c,i}$, while
their combination follows Eq.~(\ref{Eq:addition}),
provided that the individual mass-growth rates are
additive and share the $\Mbh^2$ dependence assumed in
Eq.~(\ref{Eq:massgrowthera}).


\subsection{Application: inflaton accretion and radiation absorption}

We now illustrate the use of the addition theorem by considering a
reheating era during which two distinct additive
mass-growth mechanisms act simultaneously: the absorption of the thermal
radiation bath and the accretion of the inflaton condensate. The
contribution of inflaton accretion to the PBH mass-growth rate was derived
in Ref.~\cite{Kalita:2026duh} and can be written as
\beq
\left(\frac{d\Mbh}{dt}\right)_{\rm inflaton}
=
\delta_\phi
\frac{T_{\rm RH}^4}{\Mp^4}
\Mbh^2
x^{-3(1+w_\phi)}\,,
\label{Eq:inflatonrate}
\eeq

\noi
where, consistently with the notation introduced in the previous
section, $x\equiv a/a_{\rm RH}$, and the dimensionless coefficient
$\delta_\phi$ is given by
\beq
\delta_\phi
=
\frac{\alpha}{6}
\left(\frac{1+w_\phi}{1-w_\phi}\right)^2
\left[
\frac{
\Gamma\!\left(\frac{1}{1+w_\phi}\right)
}{
\Gamma\!\left(\frac{1-w_\phi}{2(1+w_\phi)}\right)
}
{\cal P}_1^{\left(\frac{1+w_\phi}{1-w_\phi}\right)}
\right]^2\,,
\eeq

\noi
where ${\cal P}_1^{(n)}$ denotes the $n$-th harmonic of the inflaton
oscillations \cite{Kalita:2026duh}.
Eq.~(\ref{Eq:inflatonrate}) is immediately recognized as a particular
case of the generic mass-growth equation,
Eq.~(\ref{Eq:massgrowthera}), with the inflaton energy density given by
\beq
\rho_\phi
=
\rho_{\rm RH}\,
x^{-3(1+w_\phi)}\,.
\eeq

\noi
Using
\beq
\rho_{\rm RH}
=
\alpha T_{\rm RH}^{4}\,,
\eeq

\noi
the corresponding effective efficiency coefficient is
\beq
\xi_\phi
=
\delta_\phi
\frac{T_{\rm RH}^{4}}{\rho_{\rm RH}}
=
\frac{\delta_\phi}{\alpha}\,.
\label{Eq:xiphi}
\eeq

\noi
Inflaton accretion therefore constitutes an additional mass-growth
channel that can be incorporated directly into the general formalism.

The inflaton energy density scales as
$\rho_\phi\propto x^{-3(1+w_\phi)}$, corresponding to the generic
power-law evolution of Eq.~(\ref{Eq:rhophievolution}), with
$\chi_\phi=3(1+w_\phi)$. The corresponding mass amplification therefore
satisfies
\beq
\frac{1}{R_\phi}
=
1-\frac{\gamma}{\gamma_{c,\phi}}\,,
\label{Eq:Rphi}
\eeq

\noi
where, using Eq.~(\ref{Eq:calI}),
\beq
\frac{1}{\gamma_{c,\phi}}
=
{\cal A}_\phi
\int_{\xin}^{1}
x^{-1-\frac32(1+w_\phi)}\,dx\,,
\eeq

\noi
with, from Eq.~(\ref{Eq:calA}),
\beq
{\cal A}_\phi
=
4\pi\,
\xi_\phi\,
\frac{\rho_{\rm RH}}
{\Mp^2H_{\rm in}^2}\,
\xin^{-\frac32(1+w_\phi)}\,.
\eeq

\noi
Evaluating the integral gives
\beq
\frac{1}{\gamma_{c,\phi}}
=
\frac{8\pi\xi_\phi}{1+w_\phi}
\left(
1-\xin^{\frac32(1+w_\phi)}
\right)\,.
\label{Eq:gammacphi}
\eeq

\noi
For $w_\phi=0$, we recover the standard limit
$\gamma_{c,\phi}\simeq 0.75$ for $\xin\ll1$
\cite{Kalita:2026duh}.

During reheating, inflaton accretion and radiation absorption act
simultaneously and additively. Applying the addition
theorem, Eq.~(\ref{Eq:addition}), the corresponding combined critical
collapse efficiency is
\beq
\frac{1}{\gamma_{c,\rm tot}^{\rm RH}}
=
\frac{1}{\gamma_{c,\phi}}
+
\frac{1}{\gamma_{c,\rm abs}}\,,
\label{Eq:gammacRHcombined}
\eeq

\noi
where $\gamma_{c,\phi}$ is given by Eq.~(\ref{Eq:gammacphi}), while
$\gamma_{c,\rm abs}$ corresponds to the radiation-absorption threshold
derived in Eq.~(\ref{Eq:gammarunreh}), with the corresponding
logarithmic expression given in
Eq.~(\ref{Eq:gammarunreh_log}) when
$3+3w_\phi-8\zeta=0$.

The total mass amplification accumulated before the end of reheating
therefore reads, from Eq.~(\ref{Eq:Rgenericera}),
\beq
R_{\rm RH}
=
\left(
1-\frac{\gamma}{\gamma_{c,\rm tot}^{\rm RH}}
\right)^{-1}\,.
\label{Eq:RRHcombined}
\eeq

\noi
Although the addition theorem determines the total mass amplification
during reheating, the subsequent evolution is governed by the effective
collapse efficiency at the onset of radiation domination. This
effective collapse efficiency is determined by the PBH mass accumulated
before the end of reheating. Requiring it to exceed the universal
radiation-dominated threshold $\gamc$, as expressed in
Eq.~(\ref{Eq:gammacritreh}), and using Eq.~(\ref{Eq:RRHcombined})
immediately yields the runaway condition
\beq
\frac{1}{\gamma}
\leq
\frac{1}{\gamma_{c,\rm tot}^{\rm RH}}
+
\frac{\xin^{\frac32(1+w_\phi)}}{\gamc}\,.
\eeq

\noi
Defining the effective critical collapse efficiency at PBH formation,
including both reheating and the subsequent radiation-dominated era, as
$\gamma_c^{\rm RH\rightarrow RD}$, one obtains
\beq
\boxed{
\frac{1}{\gamma_c^{\rm RH\rightarrow RD}}
=
\frac{1}{\gamma_{c,\phi}}
+
\frac{1}{\gamma_{c,\rm abs}}
+
\frac{
\xin^{\frac32(1+w_\phi)}
}{
\gamc
}
}\,.
\label{Eq:fullcombinedcritical}
\eeq

As a consistency check, consider the limit
$\xin\rightarrow1$, corresponding to PBHs forming at the end of
reheating. In this limit, both reheating contributions vanish,
$\gamma_{c,\phi}^{-1}\rightarrow0$ and
$\gamma_{c,\rm abs}^{-1}\rightarrow0$, since neither inflaton accretion
nor radiation absorption has sufficient time to modify the PBH mass
before the onset of radiation domination. Eq.~(\ref{Eq:fullcombinedcritical})
therefore reduces to
\beq
\gamma_c^{\rm RH\rightarrow RD}
\longrightarrow
\gamc\,,
\eeq

\noi
thereby recovering the universal critical collapse efficiency of a
radiation-dominated Universe derived in Ref.~\cite{Haque:2026vvp}.

Eq.~(\ref{Eq:fullcombinedcritical}) constitutes the central result of
this section. It shows that the inverse critical
collapse efficiency receives additive contributions from inflaton
accretion and radiation absorption during reheating, together with the
subsequent radiation-dominated runaway condition mapped back to the PBH
formation epoch. The first two terms describe the mass growth accumulated
during reheating, whereas the last term encodes the universal
radiation-dominated threshold after reheating. This illustrates how the
addition theorem naturally combines
distinct mass-growth mechanisms acting during
reheating, while the transition to radiation domination is encoded in the
effective collapse efficiency through Eq.~(\ref{Eq:gammacritreh}).

\begin{figure}
    \centering
    \includegraphics[width=\linewidth]{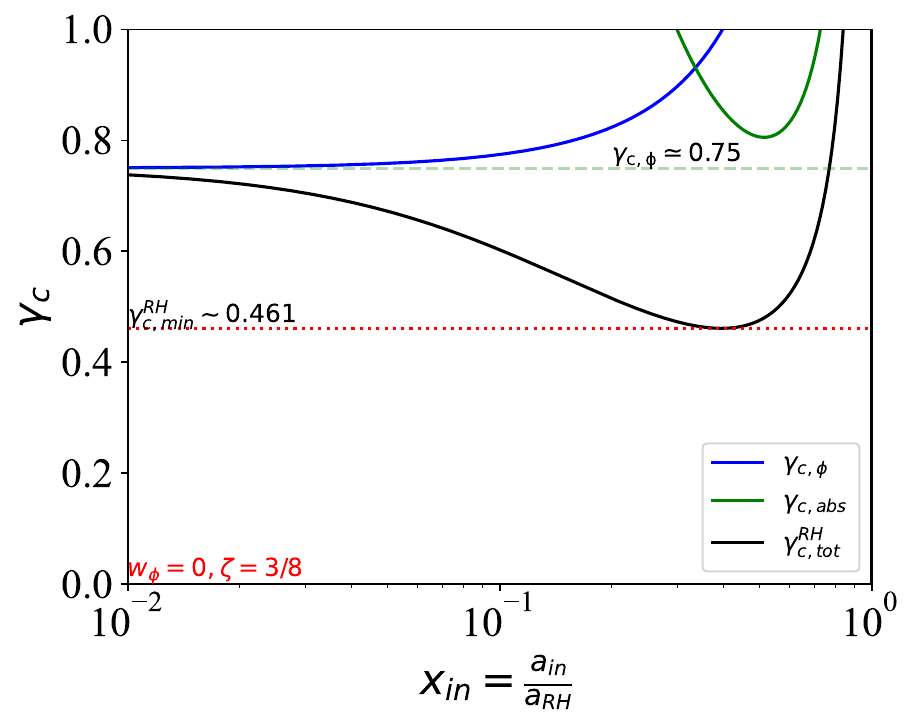}
    \caption{\justifying Critical gamma as a function of $x_{in}$ for radiation absorption (green) and inflaton accretion (blue) along with their combined contribution (black) using Eq.~\eqref{Eq:addition}.}
    \label{fig:reheating combination}
\end{figure}

\begin{figure}
    \centering
    \includegraphics[width=\linewidth]{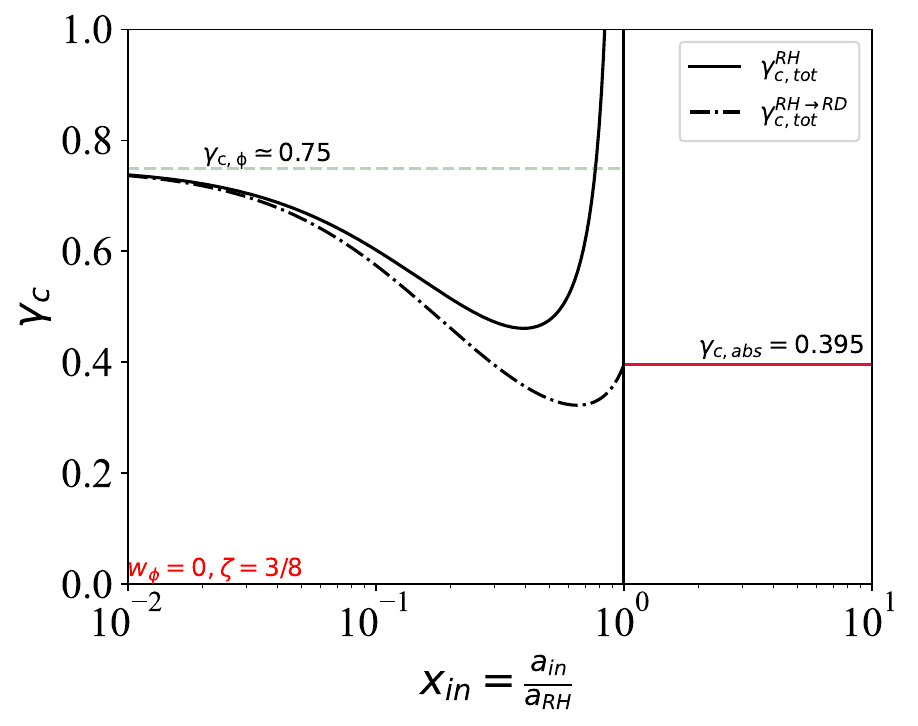}
    \caption{\justifying Critical gamma as a function of $x_{in}$ for the combined contribution during reheating (full line) and including the potential runaway during radiation domination (dash dotted line).}
    \label{fig:connected contribution}
\end{figure}

We show in Figs.~\ref{fig:reheating combination} and
\ref{fig:connected contribution} the corresponding critical collapse
efficiencies obtained by including, respectively, the combined effects of
radiation absorption and inflaton accretion during reheating, and the
additional contribution from the subsequent radiation-dominated era.
Figure~\ref{fig:reheating combination} shows that combining the two
reheating processes through the addition theorem lowers the critical
collapse efficiency to a minimum value of
$\gamma_{c,\rm min}\simeq0.461$, significantly below the threshold
associated with either process taken individually. Including the
subsequent radiation-dominated evolution further reduces the effective
critical collapse efficiency, as shown in
Fig.~\ref{fig:connected contribution}.

The minimum of the latter curve lies very close to the reheating
transition, where the piecewise approximation adopted so far becomes less
accurate. Indeed, our analytical treatment assumes that the expansion is
entirely governed by the dominant energy component in each cosmological
era. While this approximation is well justified far from the transition,
it neglects order-one corrections arising from the simultaneous presence
of the inflaton and radiation near reheating. Since the critical collapse
efficiency is sensitive to such corrections, a refined treatment of the
transition is required, as we develop in the next section.

\section{Refinement near the reheating transition}
\label{Sec:refinement}

The analytical treatment developed so far relies on a piecewise
description of the cosmological background, in which the Hubble rate is
determined by the dominant energy component in each era. This
approximation provides a simple physical interpretation and correctly
captures the evolution far from the end of reheating. However, the
critical collapse efficiency obtained in the previous section develops
its most significant variation close to the transition between inflaton
and radiation domination, where neither component can consistently be
neglected.

This is particularly important because an order-one correction to the
mass-growth rate may shift the critical collapse efficiency sufficiently
to change whether a given PBH enters the runaway regime. We therefore
refine the calculation by retaining both the inflaton and radiation
energy densities in the expression for $H$,
\beq
H^2
=
\frac{\rho_\phi+\rho_R}{3\Mp^2}\,.
\label{Eq:refinedFriedmann}
\eeq

\noi
The derivation otherwise follows the same steps as in the preceding
sections. In particular, the addition theorem remains unchanged, since
it follows from the linearity of the evolution equation for $R^{-1}$.
Only the background-dependent contribution associated with each
mass-growth mechanism must be reevaluated.


\subsection{Refined background evolution}

The Hubble rate at reheating time is therefore
\beq
H_{\rm RH}
=
\sqrt{\frac{2\alpha}{3}}\,
\frac{T_{\rm RH}^2}{\Mp}\,.
\label{Eq:refinedHRH}
\eeq

\noi
Using
\beq
\rho_\phi
=
\alpha T_{\rm RH}^4
x^{-3(1+w_\phi)}
\qquad {\rm and} \qquad
\rho_R
=
\alpha T_{\rm RH}^4
x^{-4\zeta}\,,
\eeq

\noi
the Hubble rate during reheating becomes
\beq
H
=
\frac{H_{\rm RH}}{\sqrt{2}}\,
x^{-\frac32(1+w_\phi)}
\sqrt{
1+x^{3+3w_\phi-4\zeta}
}\,.
\label{Eq:generic hubble}
\eeq

\noi
For $x\ll1$, the radiation contribution is negligible and
Eq.~(\ref{Eq:generic hubble}) reduces to the inflaton-dominated
expression used previously. Close to $x=1$, however, radiation provides
an order-one contribution to the expansion rate and must be retained.

The refined Hubble rate also modifies the relation between the PBH
formation time and its initial mass. Combining Eq.~(\ref{Eq:min})
together with Eq.~(\ref{Eq:generic hubble}), one obtains
\beq
\frac{\Mbh^{\rm in}T_{\rm RH}^2}{\Mp^3}
=
4\pi\gamma
\sqrt{\frac{3}{\alpha}}\,
\frac{
\xin^{\frac32(1+w_\phi)}
}{
\sqrt{
1+\xin^{3+3w_\phi-4\zeta}
}
}\,.
\label{Eq:generic MiT2MP3}
\eeq

\noi
Compared with Eq.~(\ref{Eq:MiT2MP3}), the additional denominator
accounts for the radiation contribution to the Hubble rate at PBH
formation. We can now use this refined background to reevaluate both
radiation absorption and inflaton accretion.


\subsection{Radiation absorption}

Using Eqs.~(\ref{Eq:generic hubble}) and
(\ref{Eq:generic MiT2MP3}), the radiation-absorption equation becomes
\beq
\frac{dR^{-1}}{d\ln x}
=
-\frac{12\pi\delta_i\gamma}{\alpha}\,
\frac{
\xin^{\frac32(1+w_\phi)}
}{
\sqrt{
1+\xin^{3+3w_\phi-4\zeta}
}
}\,
\frac{
x^{\frac{3(1+w_\phi)-8\zeta}{2}}
}{
\sqrt{
1+x^{3+3w_\phi-4\zeta}
}
}\,.
\label{Eq:refinedAbsorptionDifferential}
\eeq

\noi
For $3+3w_\phi-8\zeta\neq0$, this expression can be integrated
analytically, yielding
\begin{widetext}
\bea
\frac{1}{\Rrh}
&=&
1
-\frac{12\pi\delta_i\gamma}{\alpha}
\frac{2}{3(1+w_\phi)-8\zeta}
\frac{
\xin^{\frac32(1+w_\phi)}
}{
\sqrt{
1+\xin^{3+3w_\phi-4\zeta}
}
}
\nonumber\times
\Bigg[
{}_2F_1\!\left(
\frac12,
\frac{3(1+w_\phi)-8\zeta}{6(1+w_\phi)-8\zeta},
\frac{9(1+w_\phi)-16\zeta}{6(1+w_\phi)-8\zeta},
-1
\right)
\nonumber\\
&&\qquad
-
\xin^{\frac{3+3w_\phi-8\zeta}{2}}
\,{}_2F_1\!\left(
\frac12,
\frac{3(1+w_\phi)-8\zeta}{6(1+w_\phi)-8\zeta},
\frac{9(1+w_\phi)-16\zeta}{6(1+w_\phi)-8\zeta},
-\xin^{3(1+w_\phi)-4\zeta}
\right)
\Bigg]\,.
\label{Eq:refinedAbsorptionGeneral}
\eea
\end{widetext}

\noi
Here, ${}_2F_1$ denotes the hypergeometric function. The refined
radiation-absorption critical efficiency,
$\gamma_{c,\rm abs}^{\rm ref}$, can be read directly by rewriting
Eq.~(\ref{Eq:refinedAbsorptionGeneral}) in the form
\beq
\frac{1}{\Rrh}
=
1-\frac{\gamma}{\gamma_{c,\rm abs}^{\rm ref}}\,.
\label{Eq:refinedAbsDefinition}
\eeq

\noi
In the previously logarithmic case, $3+3w_\phi-8\zeta=0$, the solution reads
\begin{widetext}
\beq
\frac{1}{\Rrh}
=
1
-\frac{12\pi\delta_i\gamma}{\alpha}
\frac{
\xin^{\frac32(1+w_\phi)}
}{
\sqrt{
1+\xin^{3+3w_\phi-4\zeta}
}
}
\frac{2}{3(1+w_\phi)-4\zeta}
\tanh^{-1}\!\left[
\frac{
\sqrt{
1+\xin^{3(1+w_\phi)-4\zeta}
}
-\sqrt{2}
}{
1-\sqrt{
2\left(
1+\xin^{3(1+w_\phi)-4\zeta}
\right)
}
}
\right]\,.
\label{Eq:refinedAbsorptionLog}
\eeq
\end{widetext}

\noi
Both expressions recover the results obtained in the
inflaton-dominated approximation for $\xin\ll1$. Their main effect is
therefore confined to PBHs forming sufficiently close to the end of
reheating.


\subsection{Inflaton accretion}

The same procedure can be applied to inflaton accretion. We focus on
the phenomenologically relevant case $w_\phi=0$, since inflaton
accretion becomes considerably less efficient for $w_\phi>0$
\cite{Kalita:2026duh}. For perturbative reheating following a quadratic
inflaton potential, one furthermore has $\zeta=3/8$.

In this case, the refined inflaton-accretion contribution can be written
as
\begin{widetext}
\bea
\frac{1}{\Rrh}
&=&
1
-\frac{8\pi\gamma\delta_{\phi}}{\alpha}
\Bigg[
1
-
\frac{
\sqrt{2}\,\xin^{3/2}
}{
\sqrt{1+\xin^{3/2}}
}
+
\frac{
\xin^{3/2}
}{
\sqrt{1+\xin^{3/2}}
}
\tanh^{-1}\!\left(
\frac{
\sqrt{2}-\sqrt{1+\xin^{3/2}}
}{
1-\sqrt{2(1+\xin^{3/2})}
}
\right)
\Bigg]\,.
\label{Eq:refinedInflatonAccretion}
\eea
\end{widetext}

\noi

It follows immediately from Eq.~(\ref{Eq:refinedInflatonAccretion}) that
the result without refinement is recovered for $\xin\ll1$, since the expression in square
brackets approaches unity.

As for radiation absorption, the refined treatment therefore leaves
the early-time behavior unchanged and only modifies the evolution close
to the reheating transition.


\subsection{Combined critical threshold and numerical validation}

Since the addition theorem is unaffected by the refinement of the
background evolution, the refined radiation-absorption and
inflaton-accretion contributions can still be combined according to
\beq
\frac{1}{\gamma^{\rm ref}_{\rm tot}(\xin)}
=
\frac{1}{\gamma_{c,\phi}^{\rm ref}(\xin)}
+
\frac{1}{\gamma_{c,\rm abs}^{\rm ref}(\xin)}\,.
\label{Eq:refinedGammaCombined}
\eeq

\noi
The total mass amplification accumulated during reheating is then
\beq
\frac{1}{\Rrh}
=
1-\frac{\gamma}{\gamma^{\rm ref}_{\rm tot}(\xin)}\,.
\label{Eq:refinedRRH}
\eeq

\noi
To determine whether the PBH subsequently enters the runaway regime
during radiation domination, one must use the refined relation between
the effective collapse efficiency and the PBH mass at the end of
reheating. From Eq.~(\ref{Eq:generic hubble}), one finds
\beq
\frac{H_{\rm RH}}{H_{\rm in}}
=
\frac{
\sqrt{2}\,
\xin^{\frac32(1+w_\phi)}
}{
\sqrt{
1+\xin^{3+3w_\phi-4\zeta}
}
}\,.
\label{Eq:refinedHubbleRatio}
\eeq

\noi
The effective collapse efficiency at the onset of radiation domination
is therefore
\beq
\gamma_{\rm eff}
=
\gamma\Rrh
\frac{
\sqrt{2}\,
\xin^{\frac32(1+w_\phi)}
}{
\sqrt{
1+\xin^{3+3w_\phi-4\zeta}
}
}\,.
\label{Eq:refinedGammaEff}
\eeq

\noi
Requiring $\gamma_{\rm eff}\geq\gamc$ and using
Eq.~(\ref{Eq:refinedRRH}) gives the refined critical condition
\beq
\gamma
\geq
\gamma_c^{\rm ref}(\xin)
=
\frac{
\gamc\,
\gamma^{\rm ref}_{\rm tot}(\xin)\,
\sqrt{
1+\xin^{3+3w_\phi-4\zeta}
}
}{
\gamc
\sqrt{
1+\xin^{3+3w_\phi-4\zeta}
}
+
\sqrt{2}\,
\gamma^{\rm ref}_{\rm tot}(\xin)\,
\xin^{\frac32(1+w_\phi)}
}\,.
\label{Eq:refinedCriticalTotal}
\eeq

\noi
Eq.~(\ref{Eq:refinedCriticalTotal}) provides the refined counterpart of
Eq.~(\ref{Eq:fullcombinedcritical}). Far from the reheating transition,
the subdominant radiation contribution becomes negligible and the
previous analytical result is recovered, up to the $\sqrt{2}$ factor coming from the redifinition of $H_{\rm RH}$ in Eq.~(\ref{Eq:generic hubble}). Close to $\xin=1$, by contrast,
the exact two-component Hubble rate removes the artificial enhancement
generated by the piecewise approximation.

\begin{figure}[t]
\centering
\includegraphics[width = \linewidth]{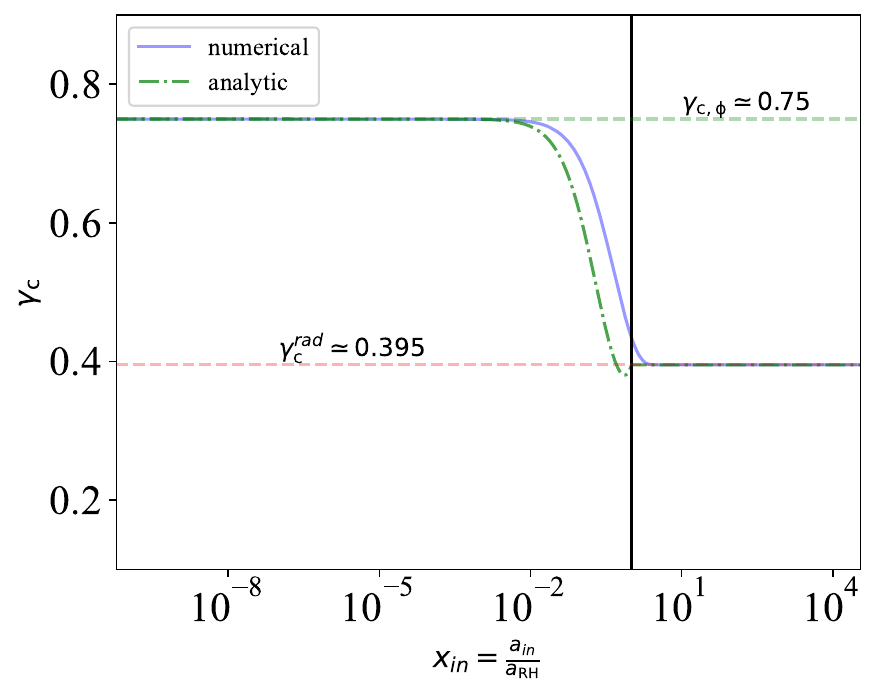}
\caption{\justifying Refined critical collapse efficiency as a function of the PBH
formation time $\xin=\ain/a_{\rm RH}$ for $w_\phi=0$ and
$\zeta=3/8$, including both inflaton accretion and radiation
absorption. The solid curve is obtained from the numerical integration
of the complete mass-evolution equation, while the dashed curve
corresponds to the refined analytical result,
Eq.~(\ref{Eq:refinedCriticalTotal}). The horizontal dashed lines mark
the asymptotic reheating-era and radiation-dominated thresholds, while
the vertical line indicates $\ain=a_{\rm RH}$.}
\label{Fig:refinedCritical}
\end{figure}

Figure~\ref{Fig:refinedCritical} compares
Eq.~(\ref{Eq:refinedCriticalTotal}) with the result obtained by
numerically integrating the complete mass-evolution equation through the
reheating-to-radiation transition. The two curves remain in excellent
agreement throughout the transition region. In particular, the refined
solution interpolates smoothly between the reheating-era threshold at
early times and the universal radiation-dominated value
$\gamc\simeq0.395$ after reheating.

Most importantly, the shallow minimum below $\gamc$ found in the
piecewise treatment almost disappears. This confirms that it is not a physical
enhancement of PBH growth, but an artifact of our approximate computations near
$\rho_\phi\simeq\rho_R$. The refined analytical treatment therefore
provides a continuous and self-consistent description of the runaway
threshold across the reheating transition.

\section{Conclusion}\label{Sec:conclusion}

The existence of a universal critical collapse efficiency during radiation domination is
a remarkable consequence of the scale-free evolution of a radiation-dominated Universe.
In this work, we have shown that this universal threshold does not generally persist when
primordial black holes form during the reheating era. The end of reheating introduces a
new physical scale, so that the critical collapse efficiency generally retains a dependence
on the PBH formation time and on the reheating dynamics, although universal asymptotic
limits may emerge in specific regimes.

We first derived analytical expressions describing the PBH mass growth during reheating
for general inflaton equations of state and temperature evolutions.
This allowed us to identify two distinct physical situations: PBHs entering the runaway regime
during reheating itself, and PBHs remaining perturbative during reheating but reaching
the universal runaway threshold only after the onset of radiation domination. We showed
that the latter criterion is the relevant one for determining the subsequent
PBH evolution, since the mass accumulated during reheating determines the effective
collapse efficiency at the beginning of the radiation era.

More generally, we established a simple composition law governing
the combined effect of several additive mass-growth mechanisms within
a given cosmological era, together with a matching prescription connecting successive
cosmological epochs. The addition theorem presented in
Eq.~(\ref{Eq:addition}) shows that the inverse critical collapse efficiencies
simply add when the corresponding mass-growth rates contribute
additively. Applying this framework to inflaton accretion and
radiation absorption during reheating, we showed how the mass accumulated through both
processes modifies the effective critical condition at the onset of radiation domination.

Finally, we refined the analytical treatment by including simultaneously the inflaton and
radiation energy densities in the Hubble expansion around the reheating transition. The
refined solution removes the artificial feature generated by the piecewise approximation
near the transition while remaining in excellent agreement with the
full numerical integration of the evolution equations. In particular,
the spurious minimum below the radiation-dominated threshold found in the piecewise
treatment disappears once both components are consistently included in the background
evolution. This demonstrates that the analytical formalism developed here provides an
accurate and self-consistent description of PBH growth
across the reheating-to-radiation transition. We also examine in Appendix~\ref{app:dof} the impact of the
temperature dependence of the relativistic degrees of freedom on the
critical collapse efficiency during radiation domination. We find that,
within the high-frequency treatment adopted here, this effect is very
small and leaves the critical value close to $\gamma_c\simeq0.395$ except for the abrupt drop of $g_{\ast}$ around the QCD transition where the threshold drops to $\gamma_c\simeq0.36$.

The formalism presented in this work can be readily
extended, in principle, to more elaborate
post-inflationary histories involving several successive reheating stages or additional
matter-dominated epochs, as well as to additional PBH mass-growth
mechanisms satisfying the assumptions of the present framework. It therefore provides a
simple analytical framework for studying PBH evolution in realistic cosmological scenarios
beyond the standard radiation-dominated picture.
\section*{Acknowledgment}

The authors would like to thank Sébastien Clesse, Guillem Domènech, Chul-Moon Yoo, Benjamin Lehmann and Miguel Vanvlasselaer for
helpful discussions. 
This project has received funding from the European Union’s Horizon Europe research and innovation programme under the Marie Skłodowska-Curie Staff Exchange grant agreement No 101086085 – ASYMMETRY and the CNRS-IRP project UCMN.
We also acknowledge support by Institut Pascal at Université 
Paris-Saclay during the Paris-Saclay Astroparticle Symposium 2025.  M.R.H. acknowledges the Tsung-Dao Lee Institute at Shanghai Jiao Tong University for financial support through the Siyuan Postdoctoral Fellowship.

\appendix

\begin{figure}
    \centering
    \includegraphics[width=\linewidth]{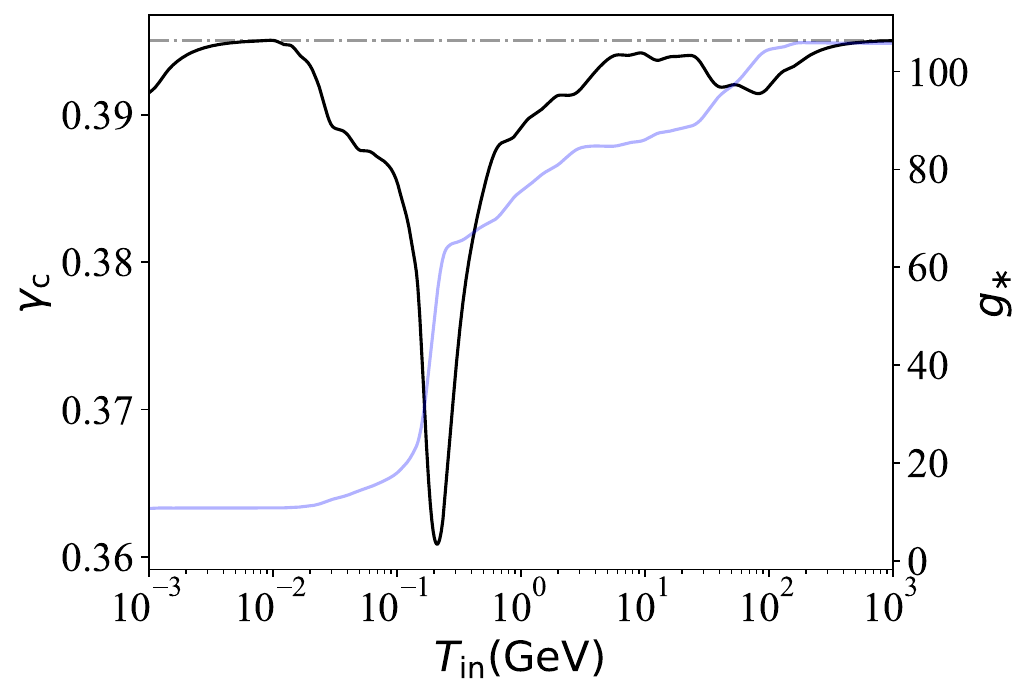}
    \caption{\justifying Critical collapse efficiency $\gamma_c$ during radiation domination
as a function of the PBH formation temperature $T_{\rm in}$. The black
curve shows $\gamma_c$ obtained by including the temperature dependence
of the relativistic degrees of freedom, while the blue curve shows
$g_*(T_{\rm in})$. The gray dot-dashed line denotes the constant-$g_*$
high-frequency result, $\gamma_{c,\rm rad}\simeq 0.395$.}
    \label{fig:d.o.f}
\end{figure}

\section{Impact of relativistic degrees of freedom on the critical collapse efficiency}
\label{app:dof}

In this Appendix, we investigate the impact of the temperature dependence
of the relativistic degrees of freedom on the critical collapse efficiency
during radiation domination. A related analysis was performed in
Ref.~\cite{Kallifatides:2026sik}, where radiative absorption was described
using the Stefan--Boltzmann law together with the so-called
``principle of isonomy.'' Since their prescription for the particle-species
dependence of the absorption rate differs from the high-frequency
geometrical-optics treatment adopted here, it is useful to examine
explicitly how the variation of the relativistic degrees of freedom affects
our critical threshold.

Starting again from Eq.~\eqref{Eq:evolution} and focusing on the
absorption-dominated regime during radiation domination, entropy
conservation implies
\beq
a\,T\,g_{*s}^{1/3}(T)={\rm const.},
\eeq
and therefore
\beq
\frac{dT}{dt}
=
-\frac{H T}
{1+\frac{1}{3}\frac{d\ln g_{*s}}{d\ln T}}\,.
\label{Eq:dTdt_dof}
\eeq

\noi
Using the radiation-dominated Friedmann equation, we obtain
\beq
\frac{d(1/R)}{dT}
=
\frac{12\pi\gamma\,\delta_i(T)}
{\sqrt{\alpha(T_{\rm in})\alpha(T)}}\,
T_{\rm in}^{-2}
\left(
1+\frac{1}{3}
\frac{d\ln g_{*s}}{d\ln T}
\right)T\,.
\label{Eq:dRdT_dof}
\eeq

For simplicity, we assume that the entropy and energy relativistic
degrees of freedom are approximately equal,
\beq
g_{*s}(T)\simeq g_*(T)\,.
\label{Eq:gs_g_approx}
\eeq
Under this approximation, Eq.~\eqref{Eq:dRdT_dof} becomes
\beq
\frac{d(1/R)}{dT}
=
\frac{12\pi\gamma\,\delta_i(T)}
{\sqrt{\alpha(T_{\rm in})\alpha(T)}}\,
T_{\rm in}^{-2}
\left(
1+\frac{1}{3}
\frac{d\ln g_*}{d\ln T}
\right)T\,.
\label{Eq:dRdT_dof_approx}
\eeq
Integrating from $T_{\rm in}$ to a temperature $T<T_{\rm in}$ gives
\beq
\begin{aligned}
\frac{1}{R(T)}
={}&1+
12\pi\gamma\,T_{\rm in}^{-2}
\int_{T_{\rm in}}^{T}
\frac{\delta_i(T')\,T'\,dT'}
{\sqrt{\alpha(T_{\rm in})\alpha(T')}}
\\
&\times
\left(
1+\frac{1}{3}
\frac{d\ln g_*(T')}
{d\ln T'}
\right).
\end{aligned}
\label{Eq:R_dof}
\eeq

Since the absorption contribution is UV dominated during radiation
domination, we take the formal late-time limit $T\rightarrow0$.
The critical collapse efficiency is obtained by imposing $1/R=0$,
yielding
\beq
\begin{aligned}
\frac{1}{\gamc}
={}&
-12\pi\,T_{\rm in}^{-2}
\int_{T_{\rm in}}^{0}
\frac{\delta_i(T)\,T\,dT}
{\sqrt{\alpha(T_{\rm in})\alpha(T)}}
\\
&\times
\left(
1+\frac{1}{3}
\frac{d\ln g_*(T)}
{d\ln T}
\right).
\end{aligned}
\label{Eq:gamcdof1}
\eeq

As a consistency check, in the high-frequency regime, when $g_*$ is
taken to be constant, $\delta_{\rm hf}$ also becomes constant and the
above expression reduces to

\beq
\gamc=\frac{\alpha}{6\pi\delta_{\rm hf}}=\frac{32}{81}\simeq0.395\,.
\eeq

The integral in Eq.~\eqref{Eq:gamcdof1} can be evaluated numerically
using an interpolating function for the temperature-dependent
relativistic degrees of freedom. In the high-frequency regime considered
here, both $\delta_{\rm hf}(T)$ and $\alpha(T)$ are proportional to
$g_*(T)$. Although this leads to a partial cancellation of the explicit
$g_*$ dependence in Eq.~\eqref{Eq:gamcdof1}, a residual dependence
remains through the ratio $g_*(T)/g_*(T_{\rm in})$ and through the
entropy-conservation correction
$1+\frac{1}{3}\frac{d\ln g_*}{d\ln T}$.
Consequently, the critical collapse efficiency approaches the
constant-$g_*$ result, $\gamma_{c,\rm rad}=32/81\simeq0.395$, whenever
$g_*(T)$ varies slowly, but can deviate appreciably from this value
across temperature intervals in which the relativistic degrees of
freedom change rapidly.

This behavior is particularly relevant around the QCD transition,
where $g_*(T)$ varies rapidly with temperature. In this region, both
the explicit temperature dependence of $g_*(T)$ and the
entropy-conservation correction proportional to
$d\ln g_*/d\ln T$ contribute to the departure from the
constant $g_*$ result. The latter becomes particularly important
across the transition and enhances the reduction of
$\gamma_c(T_{\rm in})$ visible in Fig.~\ref{fig:d.o.f}. The critical
collapse efficiency reaches a minimum of approximately
\beq
\gamma_{c,\rm min}\simeq 0.36\,,
\eeq
\noi
around the QCD transition, corresponding to a reduction of about
$10\%$ with respect to the constant-$g_*$ value, before increasing
again toward
$\gamma_{c,\rm rad}=32/81\simeq0.395$ at higher temperatures.
Away from temperature intervals where $g_*(T)$ varies rapidly, the
derivative contribution becomes small and the constant-$g_*$ result
is recovered to very good accuracy.

The resulting temperature dependence is therefore considerably larger
than would be inferred from the partial cancellation of the explicit
$g_*$ dependence alone. Interestingly,
Ref.~\cite{Kallifatides:2026sik} also finds a substantial temperature
dependence of the critical collapse efficiency, although
their treatment of the species-dependent absorption rate differs from
the high-frequency geometrical-optics prescription adopted here.







\bibliographystyle{plain}
\bibliographystyle{unsrt}
\bibliography{refs}

\end{document}